\documentclass[english,12pt,a4paper]{article}

\usepackage{graphics}
\usepackage{graphicx}
\usepackage{texdraw}
\usepackage{epsfig}
\usepackage{float}
\usepackage{amsmath}
\usepackage{amssymb}
\usepackage{enumerate}
\usepackage[sort&compress,numbers]{natbib}
\usepackage[all]{xy}
\usepackage{slashed}
\newcommand{\be}{\begin{equation}}
\newcommand{\ee}{\end{equation}}

\newcommand{\f}[2]{\frac{#1}{#2}}
\newcommand{\la}{\langle}
\newcommand{\ra}{\rangle}
\newcommand{\D}{{\cal D}}
\newcommand{\Det}{{\rm Det}\,}
\newcommand{\tr}{{\rm tr}}

\newcommand{\Tr}{{\rm Tr}}

\date{\today}

\begin{document}

\title{Phase structure of a generalized Nambu Jona-Lasinio model with Wilson 
fermions in the mean field or large $N$-expansion}

\maketitle
\setcounter{page}{1}

\renewcommand{\thefootnote}{\roman{footnote}}

\author{V. Azcoiti\footnote[1]{Departamento de F\'{\i}sica Te\'orica, Facultad de Ciencias, Universidad de Zaragoza, Cl. Pedro Cerbuna 12, E-50009 Zaragoza (Spain)},
G. Di Carlo\footnote[2]{INFN, Laboratori Nazionali del Gran Sasso, I-67100 Assergi, L'Aquila (Italy)},
E. Follana\footnotemark[1],
M. Giordano\footnote[3]{Institute of Nuclear Research of the Hungarian Academy of Sciences (ATOMKI), Bemt\'er 18/c, H-4026 Debrecen, Hungary} and
A. Vaquero\footnote[4]{Computation-based Science and Technology Research Center (CaSToRC), The Cyprus Institute, 20 Constantinou Kavafi Street, Nicosia 2121, (Cyprus)}}

\renewcommand{\thefootnote}{\arabic{footnote}}

\abstract{We analyze the vacuum structure of a generalized 
lattice Nambu--Jona-Lasinio model with two flavors of Wilson fermions,
such that its continuum action is the most general four-fermion action 
with ``trivial'' color interactions, and 
having a $SU(2)_V \times SU(2)_A$ symmetry in the chiral limit. The  
phase structure of this model in the space of the two four-fermion
couplings shows, in addition to the standard Aoki phases, new phases with 
$\la\bar\psi \gamma_5\psi\ra\ne 0$, in close analogy to similar results
recently suggested by some of us for lattice $QCD$ with two degenerate
Wilson fermions. This result shows how the phase structure of an 
effective model for low energy $QCD$ cannot be entirely understood
from Wilson Chiral Perturbation Theory, based on the standard $QCD$
chiral effective Lagrangian approach.}

\section{Introduction}
Since the first numerical investigations of four-dimensional
non-abelian gauge theories with dynamical Wilson fermions were
performed in the early 80's \cite{an,hamber}, the understanding of the
phase and vacuum structure of lattice QCD with Wilson fermions at
non-zero lattice spacing, and of the way in which chiral symmetry is
recovered in the continuum limit, has been a goal of lattice field
theorists. The 
complexity of the phase structure of this model has been known for a
long time. The existence of a phase with parity and flavor symmetry
breaking was conjectured for this model by Aoki in the middle 80's
\cite{a1,a2}, and confirmed later on 
\cite{aokinjl,creutz,bit,aku,heller,heller2,sharpe,a3,bit2,shasi,kps,
gosha,ilg,ster,golt,ilg3,monos,sharpe2,POS1,POS2,POS3,verba,Splittorff}. 

The standard wisdom on lattice $QCD$ with Wilson fermions is that even if 
chiral symmetry is explicitly broken at finite lattice spacing $a$ by the 
Wilson regularization, this symmetry will be recovered and
spontaneously broken in the continuum limit. However, it is difficult
to understand why there exists a critical line at finite lattice
spacing along which the three pions are massless. 
Indeed, the pions cannot be the three Goldstone bosons associated with the 
spontaneous breaking of the $SU(2)$ chiral symmetry since, as 
previously stated, the Wilson regularization breaks explicitly this
symmetry. 

One of the main features of Aoki's picture was to clarify this point.
In the Aoki phase, the charged pions are massless because they are the
two Goldstone bosons associated with the spontaneous breaking of the
$SU(2)$ flavor symmetry down to $U(1)$, with a non vanishing vacuum
expectation value of the $i\bar\psi\gamma_5\tau_3\psi$ condensate. The
neutral pion, which is massive in the Aoki phase, becomes massless on
the critical line because flavour symmetry is continuously recovered on 
this line which separates the broken (Aoki) phase 
from the unbroken (physical)
phase. The other relevant feature of the Aoki scenario is that it
provides a counterexample to the Vafa-Witten theorem on the
impossibility to spontaneously break parity in a vector-like theory
with positive definite integration measure \cite{witten,monos2,ji,kr,monos3}.  

Aoki's conjecture has been supported not only by numerical simulations
of lattice $QCD$, but also by theoretical studies based on the
Nambu--Jona-Lasinio model \cite{bitar,aokinjl}, on the linear sigma model
\cite{creutz}, and on applying Wilson chiral perturbation theory ($W\chi PT$)
to the continuum effective Lagrangian \cite{sharpe}.  
The latter analysis predicts, near the continuum limit,
two possible scenarios, depending on the sign of an unknown low-energy 
coefficient. In the first scenario, flavor and parity are spontaneously 
broken, and there is an Aoki phase with a nonzero value only for the
$i\bar\psi\gamma_5\tau_3\psi$ condensate, whereas 
$\la i\bar\psi\gamma_5\psi\ra = 0$. In the other one (the
``first-order'' scenario) there is no spontaneous symmetry breaking.

This standard picture for the Aoki phase was questioned by three of us 
in \cite{monos}, where we conjectured on the appearance of new vacua
in the Aoki phase, which can be characterized by a non-vanishing
vacuum expectation value of the flavor-singlet pseudoscalar condensate
$i\bar\psi\gamma_5\psi$, and which cannot be connected to the Aoki
vacua by parity-flavor symmetry transformations. More recently, we have
obtained results from numerical simulations of lattice $QCD$ with two
degenerate flavors of Wilson fermions, suggesting that our conjecture
could be realized \cite{monos4}. Since these results seem to question
the validity of the $W\chi PT$ analysis \cite{sharpe}, an approach
which has been successfully applied in many contexts, it is worthwhile
to analyze the possible origins of this discrepancy.

First, one should notice that the chiral effective Lagrangian approach 
is based
on the continuum effective Lagrangian written as a series of
contributions proportional to powers of the lattice spacing $a$, plus
the construction of the corresponding chiral effective Lagrangian,
keeping only the terms up to order $a^2$ \cite{sharpe}. This means
that predictions based on this chiral effective Lagrangian approach 
should work
close enough to the continuum limit, where keeping terms only up to
order $a^2$ can be justified. However, the data reported  
in \cite{monos4} were obtained at $\beta=2.0$, and a very rough
estimate gives a lattice spacing of order $3.0\, {\rm GeV}^{-1}$ at this
$\beta$. Hence, a possible explanation for the discrepancies that we
found relies on the necessity of including higher-order terms in the
chiral effective Lagrangian. 

We want to recall here that the Aoki effective potential was obtained 
from a strong coupling expansion combined with a $1/N$ expansion
\cite{a2}, i.e., far away from the continuum limit. Furthermore, Aoki's
solution shows degenerate vacua with 
$\la i\bar\psi\gamma_5\tau_3\psi \ra \ne 0$, $\la
i\bar\psi\gamma_5\psi \ra  = 0$ and  
$\la i\bar\psi\gamma_5\tau_3\psi \ra  = 0$, $\la i\bar\psi\gamma_5\psi
\ra \ne 0$ respectively in the strong coupling limit. The inclusion of higher
order contributions to the strong coupling-$1/N$ expansions breaks the vacuum
degeneracy by selecting the standard vacuum with $\la
i\bar\psi\gamma_5\tau_3\psi \ra \ne 0$, but the normalized difference of vacuum
energy densities of the two vacua is of order $10^{-14}$ at $\beta=2.0$ and 
$N=3$, showing the extremely high instability of the Aoki solution.

The second possible origin of the discrepancies between the non-standard 
scenario of \cite{monos,monos4} and the $W\chi PT$ analysis 
of \cite{sharpe}, the analysis of which will be the main subject of
this paper, lies in the following point. The chiral effective
Lagrangian approach is based, as it is well known, on the assumption
that the relevant low-energy degrees of freedom in $QCD$ are the three
pions. This assumption can be reliable in the physical phase, up to 
the critical line, and also in the Aoki phase, near the critical line,
but it could break down as we go deep in the Aoki phase, where the
neutral pion is massive. 
Indeed, $QCD$ with two degenerate flavors of Wilson fermions of bare
mass $m_0 = -4.0$ in lattice units should also show degenerate vacua
with $\la i\bar\psi\gamma_5\tau_3\psi \ra \ne 0$, $\la 
i\bar\psi\gamma_5\psi \ra  = 0$ and $\la i\bar\psi\gamma_5\tau_3\psi
\ra  = 0$, $\la i\bar\psi\gamma_5\psi \ra \ne 0$ respectively, as
discussed in \cite{monos4}.  

With the purpose of establishing the range of applicability of the 
standard $QCD$ chiral effective Lagrangian approach, we will analyze 
in this paper the vacuum structure of a generalized
Nambu--Jona-Lasinio model ($NJL$) with Wilson fermions in the mean
field or leading order $1/N$-expansion. The model has been chosen to
possess the more general $SU(2)_V \times SU(2)_A$ symmetry in the
continuum, in analogy to $QCD$. The election of the $NJL$ model for
our analysis was motivated by the fact that four-dimensional models
without gauge fields, and with four fermion interactions, are
considered as effective models to describe the low energy physics of
$QCD$\footnote{For a review on the NJL model, see \cite{ref2} and
references therein.} \cite{shankar,ref1}. 

The outline of the paper is as follows. In section 2 we describe the model 
in the continuum and its lattice regularized version with Wilson fermions, 
as well as the way in which the model can be analytically solved in the 
mean field--$1/N$-expansion with the help of eight auxiliary scalar and 
pseudoscalar fields. The gap equations and the phase diagram of the 
mean-field model in the various physically relevant cases are analyzed
in section 3. In section 4 we show how the mean field equations of our
generalized $NJL$ model can be obtained in the leading order of the
$1/N$-expansion of a four-fermion model with non trivial color and
flavor interactions, but where the action is local and free from the
sign problem. Section 5 summarizes our conclusions.

\section{The model}

The most general four-fermion continuum Lagrangian in Euclidean space 
with $SU(2)_V \times SU(2)_A$ symmetry in the chiral limit and with
trivial color dependence can be written as follows,
\be
-{\cal L} = - \bar\psi\left(\slashed{\partial} + m \right) \psi +
            G_1 \left[ \left(\bar\psi\psi\right)^2 + \left(i \bar\psi
            \gamma_5{\vec{\tau}}  \psi\right)^2 \right] +
            G_2 \left[ \left(i \bar\psi \gamma_5\psi\right)^2 + \left(\bar\psi
            {\vec{\tau}}  \psi\right)^2 \right]\,,
\label{NJL}
\ee
where $\psi$ is a fermion field with four Dirac and two flavor 
components, and $\tau^a$ are the Pauli matrices acting in flavor
space. It is customary, in order to avoid the sign problem and/or to
perform a $1/N$-expansion, to add another (``color'')
degree of freedom to the spinors, and to straightforwardly generalize
the interaction by replacing $\bar\psi B \psi \to
\sum_{i=1}^N\bar\psi_i B \psi_i$, where $B$ is any of the matrices
appearing in Eq.~\eqref{NJL}. Although the interaction is not diagonal
in color space, it will become so after a Hubbard-Stratonovich
transformation, and moreover it will be the same for every color: for
this reason we will call it diagonal and trivial in color, with a
small abuse of terminology. In the following, we will refer to this
straightforward generalization as the $N$-color model.

The $NJL$ model given by action \eqref{NJL} enjoys the
same $SU(2)_V \times SU(2)_A$ symmetry of $QCD$ and it is an effective model
to describe the low energy physics of $QCD$ \cite{shankar}. 
This model, regularized on a hypercubic four-dimensional lattice with
Wilson fermions, was analyzed in the $G_2=0$ limit and in 
the mean field or first
order $1/N$-expansion by Aoki et
al. \cite{aokinjl}, who found a phase, for large values of $G_1$, in
which both flavor symmetry and parity are spontaneously broken, in
close analogy to lattice $QCD$ with Wilson fermions. 
The qualitative results of 
Aoki et al. were also corroborated by Bitar and Vranas in \cite{bitar}, where 
they found, using numerical simulations, the existence of this parity-flavor 
broken phase in the two-color model. 

The lattice action of the $N$-color model in the Wilson regularization
can be written as $S=S_0+S_I$, with the free part of the action being
\begin{equation}
  \label{eq:free_action}
  S_0 = \sum_{x,y} \bar\psi_x \Delta_{xy} \psi_y\,,
\end{equation}
where now $\psi$ is a fermion field with four Dirac, two flavor and
$N$ color components, and
where the Dirac-Wilson operator $\Delta$ is given by 
\begin{equation}
  \label{eq:wilson_f}
  \Delta_{xy} =
  \f{1}{2}\sum_{\mu=1}^4\left[(\gamma_\mu-r)\delta_{x+\hat\mu,y}
- (\gamma_\mu+r)\delta_{x-\hat\mu,y}\right] + (4r + m_0)\delta_{xy}\,,
\end{equation}
with $r$ the Wilson parameter and $m_0$ the bare fermion
mass. The interaction part is
\begin{equation}
  \label{eq:interaction}
  -S_I =
 \sum_{x} \f{G_1}{N}\left[\left(\bar\psi_x\psi_x\right)^2
+\left(\bar\psi_xi\gamma_5\vec\tau\psi_x\right)^2\right] 
+\f{G_2}{N}\left[\left(\bar\psi_xi\gamma_5\psi_x\right)^2
+ \left(\bar\psi_x\vec\tau\psi_x\right)^2\right]\,,
\end{equation}
were we have conveniently redefined the coupling constants. 
As it is well known, the Wilson term breaks explicitly the full
chiral symmetry, and so only parity and vector symmetries are kept in
the lattice regularization. 
The four-fermion action can be bilinearized by performing a 
Hubbard-Stratonovich transformation, which implies the introduction of 
eight scalar and pseudoscalar auxiliary fields as follows,
\begin{equation}
  \label{eq:HS}
  S_B =
  N\sum_x[\beta_1(\sigma_x^2+\vec\pi_x^2)+\beta_2(\eta_x^2+\vec\rho_x^{\,2})] 
+\sum_{x,y} \bar\psi_x M_{xy} \psi_y\,,
\end{equation}
where the fermion matrix $M$ is
\begin{equation}
  \label{eq:M}
  M_{xy}=\Delta_{xy} + \delta_{xy}(\sigma_x + i\gamma_5\vec\tau \cdot\vec\pi_x
  +  i\gamma_5 \eta_x + \vec\tau\cdot\vec\rho_x)\,,
\end{equation}
and moreover $\beta_i=1/(4G_i)$. Here we are considering the case
$G_1,G_2\ge 0$. 

In the $G_2 = 0$ case analyzed in \cite{bitar,aokinjl} it is easy 
to see that the fermion 
determinant is real\footnote{Reality is readily proved by noting that
  $C M C^\dag= M^*$, with $C=\tau_2\gamma^1\gamma^3$ and $CC^\dag=1$.} and
therefore the theory, with an even number of colors,  
is free from the sign problem. This allows to consistently perform the 
$1/N$-expansion \cite{aokinjl} and the numerical simulations in the two-color 
model \cite{bitar}. Unfortunately, in the general case ($G_1\ne 0, G_2\ne 0$) 
the fermion determinant is complex, and even if the leading order of the 
$1/N$-expansion is free from the sign problem for an even number of colors, 
the very consistency of this expansion is, at least, doubtful. This is the 
reason why we decided to study the infinite range model, or mean field 
approximation, where again one can easily show that the sign problem
is absent for an even number of colors. However, in section 4 we will
show how the gap equations of the infinite range model are just the
same obtained at leading order in the $1/N$-expansion of a
four-fermion model with local interactions, the same symmetries, and
free from the sign problem.  

\section{The phase diagram of the mean-field model}

The interaction part $S_I^{\rm (MF)}$ of the lattice action $S^{\rm
  (MF)} = S_0+S_I^{\rm (MF)}$ for the infinite-range model can be 
written as follows, 
\begin{multline}
  \label{eq:interaction_mf}
  -S_I^{\rm (MF)} = \f{G_1}{N}\f{1}{V}\left[\left(\sum_{x}\bar\psi_x\psi_x\right)^2
+
\left(\sum_{x}\bar\psi_xi\gamma_5\vec\tau\psi_x\right)^2\right] \\
+\f{G_2}{N}\f{1}{V}\left[\left(\sum_{x}\bar\psi_xi\gamma_5\psi_x\right)^2
+ \left(\sum_{x}\bar\psi_x\vec\tau\psi_x\right)^2\right]\,,
\end{multline}
where $V$ is the number of lattice sites. Performing again a 
Hubbard-Stratonovich transformation we get for the bilinearized action
\begin{equation}
  \label{eq:HS_mf}
  S_B^{\rm (MF)} =
  VN[\beta_1(\sigma^2+\vec\pi^2)+\beta_2(\eta^2+\vec\rho^{\,2})]
+\sum_{x,y} \bar\psi_x M_{xy} \psi_y\,,
\end{equation}
where the auxiliary fields are now constant fields, the fermion matrix is
\begin{equation}
  \label{eq:M2}
  M_{xy}=\Delta_{xy} + \delta_{xy}(\sigma + i\gamma_5\vec\tau \cdot\vec\pi
  +  i\gamma_5 \eta + \vec\tau\cdot\vec\rho)\,,
\end{equation}
and again $\beta_i=1/(4G_i)$, with $G_1,G_2\ge 0$. The integral over
the fermion fields can 
again be done analytically and in the limit of large volume $V$ the
model can be solved by writing down and solving the saddle-point
equations. 

Integrating out the fermionic degrees of freedom, the partition
function of the mean-field model reads 
\begin{equation}
  \label{eq:partfunc2}
Z = \int d\sigma d^3\!\pi d\eta d^3\!\rho
\,\Det
M\,e^{-NV[\beta_1(\sigma^2+\vec\pi^2)+\beta_2(\eta^2+\vec\rho^{\,2})]}
\equiv
\int d\sigma d^3\!\pi d\eta d^3\!\rho\,e^{-2NV {\cal V}_{\rm eff}}\,,
\end{equation}
with ${\cal V}_{\rm eff}$ the effective potential per flavor and
color. As we show in the appendix, the fermionic determinant is real 
in this case; since we are taking an even number of colors, $\Det M$
is also positive, so that there is no sign problem, and we can write
$\Det M = (\Det M M^\dag)^{\f{1}{2}}$. 

In order to compute the determinant it is convenient to go over to
momentum space. Starting from a finite lattice with periodic boundary
conditions and then taking the limit of infinite volume, one obtains
\begin{equation}
  \label{eq:det1}
  \Det M = \exp\left\{ \f{V}{2} \int_B\f{d^4p}{(2\pi)^4}\, \tr \log \tilde
    M(p)\tilde M(p)^\dag + {\cal O}(V^{-1})\right\}\,,
\end{equation}
where
\begin{equation}
  \label{eq:det2}
M_{xy} = \int_B\f{d^4p}{(2\pi)^4}\, e^{-ip\cdot(x-y)}\,\tilde M(p) \,,
\qquad   \tilde M(p) = \sum_x e^{ip\cdot x} \,M_{x0}\,,
\end{equation}
and where $B$ is the first Brillouin zone $p_\mu\in[0,2\pi]$,
$\mu=1,\ldots,4$ (or equivalently $p_\mu\in[-\pi,\pi]$ due to
periodicity), and $\tr$ stands for the trace over Dirac, flavor and
color indices. A straightforward calculation shows that
\begin{equation}
  \label{eq:tildeM}
  \tilde M(p) = i\sum_{\mu=1}^4\gamma_\mu \sin p_\mu +
  r\left(4-\sum_{\mu=1}^4\cos p_\mu\right) + m_0 + \sigma + 
i\gamma_5\vec\tau \cdot\vec\pi
  +  i\gamma_5 \eta + \vec\tau\cdot\vec\rho\,.
\end{equation}
The effective potential ${\cal V}_{\rm eff}$ can be computed
explicitly, and reads
\begin{equation}
  \label{eq:veff}
  {\cal V}_{\rm eff} = \f{\beta_1}{2}(\sigma^2+\vec\pi^2) +
  \f{\beta_2}{2}(\eta^2+\vec\rho^{\,2})  - \int_B\f{d^4p}{(2\pi)^4} \log Q \,,
\end{equation}
with
\begin{multline}
  \label{eq:Q}
  Q = \Sigma(p)^2
+ 2
\Sigma(p)\left[(w_r(p)+m_0+ \sigma)^2+\vec\pi^2+(\eta^2+\vec\rho^{\,2})\right]\\ 
+\left[(w_r(p)+m_0+\sigma)^2+\vec\pi^2-(\eta^2+\vec\rho^{\,2})\right]^2
+ 4 [\eta(w_r(p)+m_0+\sigma)-\vec\rho\cdot\vec\pi]^2\,,
\end{multline}
where we have set
\begin{equation}
  \label{eq:defs0}
    \Sigma(p)=\sum_{\mu=1}^4 (\sin p_\mu)^2\,, \qquad
    w_r(p) = r\left(4-\sum_{\mu=1}^4 \cos p_\mu\right)\,.
\end{equation}
Notice that $Q\ge 0$. 

In the large volume limit, the partition function will be dominated by
the contribution coming from the minimum of the effective potential,
and so it can be computed through the saddle-point technique. 
In order to look for the minimum of the effective potential, it is
convenient to reorder the terms in Eq.~\eqref{eq:Q}. 
A little algebra allows to rewrite it as
\begin{multline}
  \label{eq:Q2}
  Q = \left[\Sigma(p) + ( w_r(p) + m_0 + \sigma)^2 + \Pi^2 + \eta^2 +
    \rho^2\right]^2 \\ - 4\left[\rho( w_r(p)+m_0 + \sigma) +
    \Pi\eta\cos\theta\right]^2 - 4(\eta^2+\rho^2)\Pi^2(\sin\theta)^2\,,
\end{multline}
where we have set 
\begin{equation}
  \label{eq:defs}
  \begin{aligned}
    \Pi &= |\vec \pi|\,, \quad \rho =|\vec\rho|\,, \\ \vec \pi &
    \cdot \vec \rho = \Pi\rho\cos\theta\,.
  \end{aligned}
\end{equation}
Setting also
\begin{equation}
  \label{eq:defs2}
  \begin{aligned}
    Q_0 &= \left[\Sigma(p) + ( w_r(p) + m_0+ \sigma)^2 + \Pi^2 + \eta^2 +
      \rho^2\right]^2 \ge 0\,,\\
    Q_1 &= 4\left[\rho(w_r(p) + m_0+\sigma) +
    \Pi\eta\cos\theta\right]^2 + 4(\eta^2+\rho^2)\Pi^2(\sin\theta)^2 \ge 0\,,
  \end{aligned}
\end{equation}
we have $Q=Q_0-Q_1$, and the effective potential can be rewritten as
\begin{equation}
  \label{eq:veff2}
{\cal V}_{\rm eff} = \f{\beta_1}{2}(\sigma^2+\Pi^2) +
  \f{\beta_2}{2}(\eta^2+\rho^{2})  - \int_B\f{d^4p}{(2\pi)^4} \left[\log Q_0
    + \log\left(1-\f{Q_1}{Q_0}\right)\right] \,.
\end{equation}
It is convenient for our purposes to group the various terms in two
different ways. The first way is
\begin{equation}
  \label{eq:veff03}
  \begin{aligned}
  {\cal V}_{\rm eff} =~& \f{\beta_1}{2}(\sigma^2+\Pi^2+\eta^2+\rho^2) 
-  \int_B\f{d^4p}{(2\pi)^4} \log Q_0 
   \\ & \phantom{cagnaccio}
   +\f{(\beta_2-\beta_1)}{2}(\eta^2+\rho^2)
  -  \int_B\f{d^4p}{(2\pi)^4} \log\left(1-\f{Q_1}{Q_0}\right)\,,
\end{aligned}
\end{equation}
which as we will see is appropriate for the case $\beta_1<\beta_2$,
and the second way is
\begin{equation}
  \label{eq:veff04}
  \begin{aligned}
{\cal V}_{\rm eff} &~= \f{\beta_1}{2}\sigma^2  +
  \f{\beta_2}{2}(\Pi^2 + \eta^2+\rho^2)
- \int_B\f{d^4p}{(2\pi)^4} \log Q_0\\ & \phantom{cagnaccio}
+\f{(\beta_1-\beta_2)}{2}\Pi^2
  - \int_B\f{d^4p}{(2\pi)^4}  \log\left(1-\f{Q_1}{Q_0}\right)\,,
  \end{aligned}
\end{equation}
which is appropriate for the case $\beta_1>\beta_2$. The key
observation is that $Q_0$ depends only on $\sigma$ and 
on the combination $z^2\equiv \Pi^2 + \eta^2+\rho^2$, so that
the first line in both equations depends only on $\sigma$ and
$z$. Therefore, the minimization of the effective potential at fixed
$\sigma$ and $z$ involves only the second line of
Eqs.~\eqref{eq:veff03} and \eqref{eq:veff04}. 

To make things more
transparent, let us introduce the new set of variables
$z,\omega,\varphi$, in terms of which one writes 
\begin{equation}
  \label{eq:defs3}
  \begin{aligned}
    \Pi  &= z\cos\omega\,,\\
    \eta &= z\sin\omega\cos\varphi\,,\\
    \rho &= z\sin\omega\sin\varphi\,.
  \end{aligned}
\end{equation}
The range of these variables is $z\ge 0$, $\omega\in [0,\f{\pi}{2}]$,
$\varphi\in[0,\pi]$, that corresponds to the range $\Pi\ge 0$, $\rho\ge
0$, $\eta\in\mathbb{R}$ of the original variables. In terms of the new
variables, Eq.~\eqref{eq:veff03} reads
\begin{equation}
  \label{eq:veff3}
  \begin{aligned}
  {\cal V}_{\rm eff}(\sigma,z,\omega,\varphi) 
 =~& \f{\beta_1}{2}(\sigma^2 + z^2) - \int_B\f{d^4p}{(2\pi)^4} \log Q_0(\sigma,z)
 \\ & \phantom{cagna} 
-\f{\Delta\beta}{2}(z\sin\omega)^2
-\int_B\f{d^4p}{(2\pi)^4}\log\left(1-
  \f{Q_1(\sigma,z,\omega,\varphi)}{Q_0(\sigma,z)}\right)   
\,,
\end{aligned}
\end{equation}
where $\Delta\beta\equiv \beta_1-\beta_2$ and we have made explicit
the dependence on the relevant variables, and Eq.~\eqref{eq:veff04}
reads 
\begin{equation}
  \label{eq:veff4}
  \begin{aligned}
{\cal V}_{\rm eff}(\sigma,z,\omega,\varphi) 
  &~= \f{\beta_1}{2}\sigma^2 + \f{\beta_2}{2} z^2 -
  \int_B\f{d^4p}{(2\pi)^4} \log 
  Q_0(\sigma,z)  \\ & \phantom{cagnacio}+
\f{\Delta\beta}{2}
(z\cos\omega)^2
-\int_B\f{d^4p}{(2\pi)^4}\log\left(1-
  \f{Q_1(\sigma,z,\omega,\varphi)}{Q_0(\sigma,z)}\right)   
 \,.
  \end{aligned}
\end{equation}
As we have already noted, the first line in Eqs.~\eqref{eq:veff3} and
\eqref{eq:veff4} depends only on $\sigma$ and $z$. As a consequence,
in order to minimize the effective potential with respect to $\omega$
and $\varphi$ we have to focus on the second line only. Moreover, 
since $Q\ge 0$, we have that $Q_1 \le Q_0$,
and so the last term in Eqs.~\eqref{eq:veff3} and \eqref{eq:veff4} is
positive or zero, 
\begin{equation}
  \label{eq:lastterm}
  \Delta{\cal V} \equiv -\int_B\f{d^4p}{(2\pi)^4}\log\left(1-
  \f{Q_1(\sigma,z,\omega,\varphi)}{Q_0(\sigma,z)}\right) \ge 0\,.
\end{equation}
Therefore, the second line in Eq.~\eqref{eq:veff3} is positive or zero
if $\beta_2\ge\beta_1$, and the second line in Eq.~\eqref{eq:veff4} is
positive or zero if $\beta_1\ge\beta_2$; in particular, both terms are
positive or zero. If we can find values of $\omega$ and $\varphi$ such
that these lower bounds are saturated, then we have automatically
minimized the effective potential with respect to $\omega$ and
$\varphi$. In order to do so, we need to make both terms vanish, and
in particular we need that $Q_1$ vanishes identically as a function of
the momentum.\footnote{In principle it is sufficient that $Q_1$ is
  nonzero only on a set of zero measure in the four-dimensional
  momentum space, but it is easy to see that either $Q_1$ vanishes
  identically or it vanishes on a three-dimensional hypersurface.} In
terms of our new variables, $Q_1$ reads
\begin{equation}
  \label{eq:sol3}
  Q_1 = 4(z\sin\omega)^2\left\{(z\cos\omega\sin\theta)^2 +
\left[\sin\varphi( w_r(p) + m_0 +\sigma) +
  z\cos\omega\cos\varphi\cos\theta\right]^2\right\}\,.
\end{equation}
One sees immediately that $Q_1$ vanishes identically if $z=0$, or if
$\sin\omega=0$, i.e., $\omega=0$. If $z\ne 0$, $\omega\ne 0$, then 
both terms in braces must be zero, and the second one must be so
independently of $p$: this can happen only if $\sin\varphi=0$, i.e.,
$\varphi=0,\pi$, which in turn requires that $\cos\omega=0$, i.e.,
$\omega=\f{\pi}{2}$. 

Summarizing, $Q_1$ vanishes identically only if\footnote{Roughly
  speaking, since at $z=0$ all values of $\omega$ and $\varphi$ 
are equivalent, these two cases include also the case $z=0$.}
\begin{equation}
  \label{eq:sol7}
  \begin{aligned}
    1. &~~ \omega = 0\,; \\
    2. &~~ \omega = \f{\pi}{2}\,,~\varphi=0,\pi\,,\\
  \end{aligned}
\end{equation}
independently of the values of $\sigma$ and $z$. Stated differently,
in terms of the original variables, $Q_1$ vanishes identically only if 
\begin{equation}
  \label{eq:sol7bis}
  \begin{aligned}
  1. &~~ \rho=0\,, \quad |\eta| = 0\,;\\
  2. &~~ \rho=0\,, \quad \Pi=0\,.
  \end{aligned}
\end{equation}
It is immediate to check that in case 1 the second line of
Eq.~\eqref{eq:veff3} vanishes, while in case 2 the second line of
Eq.~\eqref{eq:veff4} vanishes, independently of $\Delta\beta$. Let 
us now discuss the various cases separately.

\begin{figure}[t]
  \centering
  \includegraphics[width=0.7\textwidth]{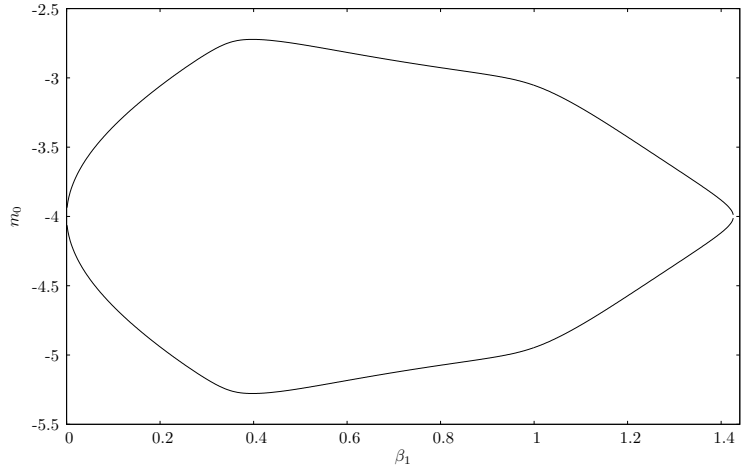}
  \caption{\footnotesize Boundary of the phase with broken parity and flavour for 
  $\Delta\beta =\beta_1-\beta_2 < 0$ in the $(\beta_1,m_0)$
  plane. When $\Delta\beta=0$, i.e., $\beta_1=\beta_2$, two degenerate
  vacua with broken parity exist, one with broken and one with
  unbroken flavour symmetry. Here we set $r=1$.}   
  \label{fig:boundary_b1}
\end{figure}

\paragraph{Case $\beta_1 <\beta_2$} In this case, the minimum of the
effective potential lies on the curve $\omega=0$, and is obtained by
minimizing the functional 
\begin{equation}
  \label{eq:final1}
  {\cal V}_< = \f{\beta_1}{2}(\sigma^2 + z^2) - \int_B\f{d^4p}{(2\pi)^4}
  \log Q_0(\sigma,z)\,, 
\end{equation}
with respect to $\sigma$ and $z$. Since $\omega=0$ corresponds to  
$\eta=\rho=0$, in this case $z=\Pi$. 
Clearly, the minimum will be independent of $\beta_2$. 
The gap equations read therefore
\begin{equation}
  \label{eq:ge1}
  \begin{aligned}
    0 =&~ \f{\beta_1}{4}\sigma - \int_B\f{d^4p}{(2\pi)^4}\,\f{w_r(p) + m_0
      +\sigma}{\Sigma(p) + (w_r(p) + m_0 +\sigma)^2 + \Pi^2}\,,\\
    0 =&~ \left[\f{\beta_1}{4} -
      \int_B\f{d^4p}{(2\pi)^4}\,\f{1}{\Sigma(p) + (w_r(p) + m_0 +\sigma)^2 +
        \Pi^2}\right]\Pi \,.
  \end{aligned}
\end{equation}
There are two solutions to these equations. The first one has $\Pi=0$
and $\sigma$ determined by the solution to the equation
\begin{equation}
  \label{eq:ge1_1}
  \begin{aligned}
   \f{\beta_1}{4}\sigma &= \int_B\f{d^4p}{(2\pi)^4}\,\f{w_r(p) + m_0
    +\sigma}{\Sigma(p) + (w_r(p) + m_0 +\sigma)^2}\,,
  \end{aligned}
\end{equation}
which always exists.\footnote{To see this it is enough to show that
  the right-hand side is always finite, and that it vanishes as
  $1/\sigma$ for large $|\sigma|$. The second point is trivial, while
  to prove the first one it is enough to bound the right-hand side as
  follows:
\begin{equation*}
    \int_B\f{d^4p}{(2\pi)^4}\,\f{w_r(p) + m_0 +\sigma}{\Sigma(p) +
      (w_r(p) + m_0 +\sigma)^2} 
\le (8 + m_0
    +\sigma)\int_B\f{d^4p}{(2\pi)^4}\,\f{1}{\Sigma(p)} < \infty\,.
  \end{equation*}
}
A second solution with nonzero $\Pi$ is obtained by solving 
\begin{equation}
  \label{eq:ge1_2}
  \begin{aligned}
0&=    \int_B\f{d^4p}{(2\pi)^4}\,\f{w_r(p) + m_0}{\Sigma(p) + (w_r(p) + m_0
      +\sigma)^2 + \Pi^2} \,,\\ 
     \f{\beta_1}{4} &=
      \int_B\f{d^4p}{(2\pi)^4}\,\f{1}{\Sigma(p) + (w_r(p) + m_0 +\sigma)^2 +
        \Pi^2} \,.
  \end{aligned}
\end{equation}
This solution is clearly degenerate since only the modulus of
$\vec\pi$ is determined, while the direction is arbitrary on the sphere
$S^2$. As it has been shown in Ref.~\cite{aokinjl}, 
a solution to
Eq.~\eqref{eq:ge1_2} exists only in a closed and bounded region in the
$(\beta_1,m_0)$ plane (see Fig.~\ref{fig:boundary_b1}); on the other
hand, when it 
exists it also minimizes the effective potential. Therefore, inside
the region enclosed in the solid line in Fig.~\ref{fig:boundary_b1},
we have a phase 
with $\Pi\ne 0$, so that $\la \bar\psi i\gamma_5\vec\tau\psi\ra =
-2N\beta_1\vec\pi\ne 0$, and thus parity and flavor are broken. On the
other hand $\la \bar\psi i\gamma_5\psi\ra = -2N\beta_2\eta =0$, so that
the vacuum is 
of the standard Aoki type.

\begin{figure}[t]
  \centering
  \includegraphics[width=0.7\textwidth]{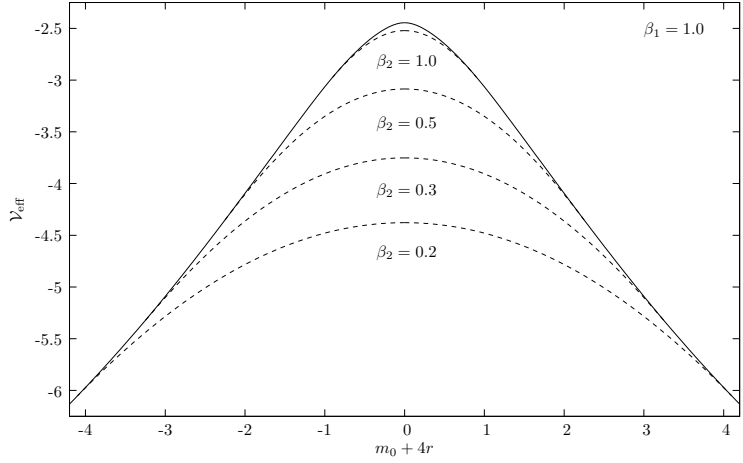}
  \caption{\footnotesize Effective potential for the solution with $\eta=0$,
    Eq.~\eqref{eq:ge1_1} (solid line), 
    and for the solution with $\eta\ne 0$,
    Eq.~\eqref{eq:ge2_2}, in the case $\beta_1=1.0$ and for various
    values of $\beta_2\le \beta_1$. Here we set $r=1$.}
  \label{fig:potential}
\end{figure}

\paragraph{Case $\beta_1 >\beta_2$} In this case, the minimum of the
effective potential lies on the curves $\omega= \f{\pi}{2}$,
$\varphi=0,\pi$, and is obtained by minimizing the functional 
\begin{equation}
  \label{eq:final2}
  {\cal V}_> = \f{\beta_1}{2}\sigma^2 + \f{\beta_2}{2} z^2 -
  \int_B\f{d^4p}{(2\pi)^4} \log Q_0(\sigma,z)
\end{equation}
with respect to $\sigma$ and $z$. Since $\omega= \f{\pi}{2}$,
$\varphi=0,\pi$, corresponds to $\Pi=\rho=0$, in this case
$z=|\eta|$. Notice that $\varphi=0$ and $\varphi=\pi$ give the same
${\cal V}_>$.  The gap equations read therefore
\begin{equation}
  \label{eq:ge2}
  \begin{aligned}
    0 =&~ \f{\beta_1}{4}\sigma - \int_B\f{d^4p}{(2\pi)^4}\,\f{w_r(p) + m_0
      +\sigma}{\Sigma(p) + (w_r(p) + m_0 +\sigma)^2 + \eta^2}\,,\\
    0 =&~ \left[\f{\beta_2}{4} -
      \int_B\f{d^4p}{(2\pi)^4}\,\f{1}{\Sigma(p) + (w_r(p) + m_0 +\sigma)^2 +
        \eta^2}\right]|\eta| \,.
  \end{aligned}  
\end{equation}
There are two solutions to these equations, one with $|\eta|=0$ and
$\sigma$ determined by the solution to Eq.~\eqref{eq:ge1_1}, 
which always exists, 
and a second solution with nonzero $|\eta|$, obtained by solving
\begin{equation}
  \label{eq:ge2_2}
  \begin{aligned}
\f{\Delta\beta}{4}\sigma &=    \int_B\f{d^4p}{(2\pi)^4}\,\f{w_r(p)
  + m_0}{\Sigma(p) + (w_r(p) + m_0 +\sigma)^2 + \eta^2} \,,\\  
     \f{\beta_2}{4} &=
      \int_B\f{d^4p}{(2\pi)^4}\,\f{1}{\Sigma(p) + (w_r(p) + m_0 +\sigma)^2 +
        \eta^2} \,,
  \end{aligned}
\end{equation}
that exists only in a certain region of the parameter space. 
This solution is (at least) twofold degenerate, since only the
absolute value of $\eta$ is determined, while the sign is
not. 

\begin{figure}[t]
  \centering
  \includegraphics[width=0.7\textwidth]{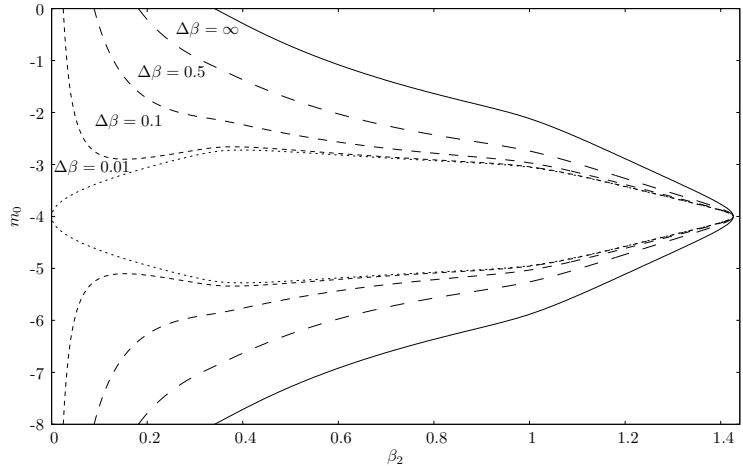}
  \caption{\footnotesize Boundary of the phase with broken parity for various finite
    values of $\Delta\beta= \beta_1-\beta_2 >0$ in the $(\beta_2,m_0)$
    plane (dashed lines). The solid line corresponds to the case
    $\Delta\beta=\infty$, i.e., $G_1=0$. The dotted line corresponds
    to the case $\Delta\beta=0$, see Fig.~\ref{fig:boundary_b1}, where
    two degenerate vacua with broken parity exist, one with broken and
    one with unbroken flavour symmetry. Here we set $r=1$.}
  \label{fig:boundary}
\end{figure}

Numerical investigations show that the solution to
Eq.~\eqref{eq:ge2_2}, when it exists, is unique (up to the sign of
$\eta$) and yields the absolute minimum of the effective potential,
see Fig.~\ref{fig:potential}. In
particular, for any choice of $\Delta\beta= \beta_1-\beta_2 >0$,
there is a region in the $(\beta_2,m_0)$ plane where the solution
exists and minimizes the potential, see Fig.~\ref{fig:boundary}. 
This region is symmetric with respect to $m_0=-4r$, and does not
extend beyond $\beta=\beta_c\simeq 1.43057$. 
This region corresponds to a phase where the condensate $\la
\bar\psi i\gamma_5\psi\ra = -2N\beta_2\eta\ne 0$, so that parity is
broken; on the other hand, since $\Pi = \rho=0$, flavor is not
broken. The vacuum is therefore not of the standard Aoki type. 
Notice that $\eta=0$ on the phase boundary.

\paragraph{Case $\beta_1 =\beta_2\equiv\beta$} In this case there are
two degenerate sets of minima, one with $\omega=0$, and one with
$\omega= \f{\pi}{2}$, $\varphi=0,\pi$, both obtained by minimizing the
functional 
\begin{equation}
  \label{eq:final3}
  {\cal V}_= = \f{\beta}{2}(\sigma^2 + z^2) - \int_B\f{d^4p}{(2\pi)^4}
  \log Q_0(\sigma,z) 
\end{equation}
with respect to $\sigma$ and $z$. The minimum at $\omega=0$
corresponds to $\eta=\rho=0$ and $z=\Pi$, while the minima at
$\omega= \f{\pi}{2}$, $\varphi=0,\pi$ correspond to $\Pi=\rho=0$ and
$z=|\eta|$. The gap equations read therefore
\begin{equation}
  \label{eq:ge3}
  \begin{aligned}
    0 =&~ \f{\beta}{4}\sigma - \int_B\f{d^4p}{(2\pi)^4}\,\f{w_r(p) + m_0
      +\sigma}{\Sigma(p) + (w_r(p) + m_0 +\sigma)^2 + z^2}\,,\\
    0 =&~ \left[\f{\beta}{4} -
      \int_B\f{d^4p}{(2\pi)^4}\,\f{1}{\Sigma(p) + (w_r(p) + m_0 +\sigma)^2 +
        z^2}\right]z \,.
  \end{aligned}
\end{equation}
There are two solutions to these equations, one with $z=0$ and
$\sigma$ determined by solving Eq.~\eqref{eq:ge1_1}, 
which can always be done, 
and a second set of solutions with nonzero $z$, obtained by solving
\begin{equation}
  \label{eq:ge3_2}
  \begin{aligned}
0&=    \int_B\f{d^4p}{(2\pi)^4}\,\f{w_r(p) + m_0}{\Sigma(p) + (w_r(p)
  + m_0 
      +\sigma)^2 + z^2} \,,\\ 
     \f{\beta}{4} &=
      \int_B\f{d^4p}{(2\pi)^4}\,\f{1}{\Sigma(p) + (w_r(p) + m_0 +\sigma)^2 +
        z^2} \,,
  \end{aligned}
\end{equation}
which is possible in a certain region of the $(\beta,m_0)$ plane. 
Setting $\eta=\rho=0$ and $z=\Pi$, there is a $S^2$ degeneracy, while
setting $\Pi=\rho=0$ and $z=|\eta|$ there is a twofold
degeneracy. Clearly, since this equation is identical to
Eq.~\eqref{eq:ge1_2}, up to the substitutions $\Pi\to z$ and
$\beta_1\to\beta$, a region exists in the $(\beta,m_0)$ plane where
the absolute minimum lies at nonzero $z$, which coincides with the one
found by Aoki and collaborators, see Fig.~\ref{fig:boundary_b1}. In this
case, however, there are two degenerate (sets of) vacua with broken
parity, one with broken and one with unbroken flavor symmetry, that
are not connected by a symmetry transformation. In this case we have
the coexistence of standard Aoki and non-standard Aoki vacua.

\section{A model with local interactions}
\label{sec:4}

In this section we describe a $N$-color model with local interactions,
free from the sign problem, which in the limit of large $N$ has the
same phase diagram of the mean-field theory, discussed in the previous
section. The action of the model is $S^{\rm (loc)}=S_0+S_I^{\rm
  (loc)}$, with the free part of the action being given in
Eqs.~\eqref{eq:free_action} and \eqref{eq:wilson_f}, and the
interaction part being
\begin{equation}
  \label{eq:loc_1}
  -S_I^{\rm (loc)} = \sum_{x} \f{G_1}{N}\left[\left(\bar\psi_x\psi_x\right)^2
+\left(\bar\psi_xi\gamma_5\vec\tau\psi_x\right)^2\right] 
+\f{G_2}{N}\left[\left(\bar\psi_xi\gamma_5\lambda\psi_x\right)^2
+ \left(\bar\psi_x\vec\tau\lambda\psi_x\right)^2\right]\,,
\end{equation}
where we are taking an even number of colors $N$, and where the
$N\times N$ diagonal matrix
$\lambda$ acting in color space is
\begin{equation}
  \label{eq:loc_2}
  \lambda = {\rm diag}(1,-1,1,-1,\ldots,1,-1)\,.
\end{equation}
After a Hubbard-Stratonovich transformation, the partition function
takes the form
\begin{equation}
  \label{eq:loc_3}
Z = \int \D\sigma \D\vec\pi \D\eta \D\vec\rho
\,\Det \bar M\,e^{-N[\sum_x\beta_1(\sigma_x^2+\vec\pi_x^2)+\beta_2(\eta_x^2+\vec\rho_x^{\,2})]}
\equiv
\int  \D\sigma \D\vec\pi \D\eta \D\vec\rho\,e^{-S_{\rm eff}}\,,
\end{equation}
where now the fermion matrix is
\begin{equation}
  \label{eq:loc_4}
\bar M_{xy}=\Delta_{xy} + \delta_{xy}(\sigma_x + i\gamma_5\vec\tau \cdot\vec\pi_x
  +  i\gamma_5 \lambda \eta_x + \lambda\vec\tau\cdot\vec\rho_x)\,.  
\end{equation}
It is easy to show that there is no sign problem in this case. Indeed, 
in color space the fermion matrix has the form
\begin{equation}
  \label{eq:loc_5}
  \begin{aligned}
  \bar M_{xy} &= {\rm
    diag}(M_{xy},M_{xy}',M_{xy},M_{xy}',\ldots,M_{xy},M_{xy}')\,, \\
   M_{xy}&=\Delta_{xy} + \delta_{xy}(\sigma_x + i\gamma_5\vec\tau \cdot\vec\pi_x
   +  i\gamma_5 \eta_x + \vec\tau\cdot\vec\rho_x)\,,\\
   M_{xy}'&=\Delta_{xy} + \delta_{xy}(\sigma_x + i\gamma_5\vec\tau \cdot\vec\pi_x
   -  i\gamma_5 \eta_x -\vec\tau\cdot\vec\rho_x)\,,
  \end{aligned}
\end{equation}
and one easily sees that $M_{xy}'=C^\dag M_{xy}^* C$, with
$C=\gamma^1\gamma^3\tau_2$. Therefore,
\begin{equation}
  \label{eq:loc_6}
  \Det \bar M = (\Det M \Det M')^{\f{N}{2}} = (\Det M \Det
  M^*)^{\f{N}{2}} = |\Det M|^N\,,
\end{equation}
which is clearly real and positive.
We can therefore write for the effective action 
\begin{equation}
  \label{eq:loc_7}
S_{\rm eff}=
\sum_x\beta_1(\sigma_x^2+\vec\pi^2)+\beta_2(\eta_x^2+\vec\rho_x^{\,2})
- \f{1}{2}(\Tr\log M + \Tr \log M^\dag)\,,
\end{equation}
where $\Tr$ is here the trace over spacetime, Dirac and flavor indices.

In the limit of large $N$, we can parameterize the auxiliary fields as
$\phi_x^i = \phi^i + \f{\delta\phi_x^i}{\sqrt{N}}$, with  
 space-independent values $\phi^i$, to be determined by the
saddle-point equations $\f{\delta {\cal V}_{\rm
    eff}}{\delta\phi_x}=0$, where we have denoted collectively 
$\{\phi_x^i\}=(\sigma_x,\vec\pi_x,\eta_x,\vec\rho_x)$. Expanding the
effective action with respect to $\delta\phi_x^i$ we find
\begin{equation}
  \label{eq:loc_8}
  S_{\rm eff}(\phi_x) = S_{\rm eff}(\phi) +
  \f{1}{2N}\sum_{x,y,i,j}\delta\phi_x^i\f{\delta^2
    S_{\rm eff}}{\delta\phi_x^i\delta\phi_y^j}\delta\phi_y^j + \ldots
\end{equation}
Going over to momentum space, and taking the limit of large volume,
we have 
\begin{equation}
  \label{eq:loc_9}
  S_{\rm eff}(\phi_x) = 2NV{\cal V}_{\rm eff}(\phi) +
  \f{1}{2}\int_B\f{d^4q}{(2\pi)^4}\, {\delta\tilde\phi}(-q)
G^{-1}_{ij}(q){\delta\tilde\phi}(q) + {\cal O}(N^{-1})\,,
\end{equation}
where we have made explicit the dependence on $N$. 
The effective potential at the saddle point reads
\begin{equation}
  \label{eq:loc_10}
  {\cal V}_{\rm eff}  = \f{\beta_1}{2}(\sigma^2+\vec\pi^2) +
  \f{\beta_2}{2}(\eta^2+\vec\rho^{\,2})  - \int_B\f{d^4p}{(2\pi)^4} \log Q \,,
\end{equation}
with $Q$ given in Eq.~\eqref{eq:Q}, 
i.e., the same result Eq.~\eqref{eq:veff} obtained in the mean-field
case. Obviously, the phase diagram at large $N$ of the local model
discussed here is therefore exactly the same found for the infinite-range model.

The first correction is expressed through the matrix
\begin{equation}
  \label{eq:loc_11}
 G^{-1}_{ij}(q) = 2\delta_{ij}\beta_{n_i} + \int_B\f{d^4p}{(2\pi)^4}
  \f{1}{g(p+\f{q}{2},z)g(p-\f{q}{2},z)}\tr\left[{\cal M}_i \tilde
    M^\dag(p+\textstyle\f{q}{2}) {\cal M}_j \tilde M^\dag(p-\textstyle\f{q}{2})
\right]\,,
\end{equation}
where $i,j=\sigma,\vec\pi,\eta,\vec\rho$, and moreover  
$n_\sigma=n_{\vec\pi}=1$, $n_\eta=n_{\vec\rho}=2$, the matrices
${\cal M}_i$ are
\begin{equation}
\label{eq:loc_12}
  {\cal M}_\sigma = 1\,, \quad
{\cal M}_{\vec\pi} = i\gamma_5\vec\tau\,,\quad
{\cal M}_\eta = i\gamma_5\,,\quad
{\cal M}_{\vec\rho}= \vec\tau\,,
\end{equation}
and the matrix $\tilde M(p)$ is given in Eq.~\eqref{eq:tildeM}. The
function $g(p,z)$ is given by 
\begin{equation}
  \label{eq:loc_13}
  g(p,z)=\Sigma(p) + (w_r(p)+m_0 +\sigma)^2 + z^2\,,
\end{equation}
with $z=0$ in the unbroken phase, $z=|\vec\pi|$ in the standard Aoki
phase, and $z=\eta$ in the non standard Aoki phase.
The nonzero elements of the inverse propagator $G^{-1}_{ij}(q)$ in the
various phases are given in the appendix. It is interesting to check
the values in the broken phases of the masses of the excitations
corresponding to $\vec\pi$ and $\eta$, defined through the small-$q$
behavior of the corresponding diagonal terms of the inverse
propagator.\footnote{In the broken phases, the $\vec\pi$'s and the
  $\eta$ are not the physical states, since they mix respectively with
  $\sigma$ and $\vec\rho$ in the standard Aoki phase, and with $\vec\rho$
  and $\sigma$ in the non standard Aoki phase.} Choosing for
convenience $\pi_a = \Pi\delta_{a3}$, we set
\begin{equation}
  \label{eq:mass}
  \begin{aligned}
    G^{-1}_{\pi_a\pi_a}(q) & \mathop =_{q\to 0} Z_{\pi_a}^{-1}(q^2 +
    m_{\pi_a}^2)\,,\\
    G^{-1}_{\eta\eta}(q)& \mathop =_{q\to 0} Z_\eta^{-1}(q^2+m_\eta^2)\,,
  \end{aligned}
\end{equation}
where $q^2=\sum_{\mu=1}^4q_\mu^2$. 
Using the results reported in the appendix, we find
that\footnote{Although we have not been able to prove it in the
  general case, the denominators in Eqs.~\eqref{eq:small_q_main} and
  \eqref{eq:small_q_main2} are certainly positive at small and at 
  large $\Pi$ or  $\eta$; furthermore, we have checked numerically
  that they are positive in the whole broken phase.}
 \begin{equation}
   \label{eq:small_q_main}
   \begin{aligned}
    m_{\pi_3}^2 &=
    \f{2\Pi^2I_2(0,\sigma,\Pi)}{I_0^{(2)}(\sigma,\Pi)-
      2\Pi^2I_2^{(2)}(\sigma,\Pi)}  \,, \qquad m_{\pi_{1,2}}^2 = 0\,,
    \\
    m_{\eta}^2 &= \f{\f{\beta_2-\beta_1}{4}+
      2\Pi^2I_2(0,\sigma,\Pi)}{I_0^{(2)}(\sigma,\Pi)- 
      2\Pi^2I_2^{(2)}(\sigma,\Pi)}\,,
   \end{aligned}
 \end{equation}
in the standard Aoki phase ($\Pi\ne 0$, $\eta=0$), and
 \begin{equation}
   \label{eq:small_q_main2}
   \begin{aligned}
    m_{\pi_a}^2 &=
    \f{\f{\beta_1-\beta_2}{4}+2\eta^2I_2(0,\sigma,\eta)}{I_0^{(2)}(\sigma,\eta)-
      2\eta^2I_2^{(2)}(\sigma,\eta)}  \,,   \phantom{\qquad
      m_{\pi_{1,2}}^2 = 0\,,}  \\ 
    m_{\eta}^2 &= \f{
      2\eta^2I_2(0,\sigma,\eta)}{I_0^{(2)}(\sigma,\eta)- 
      2\eta^2I_2^{(2)}(\sigma,\eta)}\,,
   \end{aligned}
 \end{equation}
in the non standard Aoki phase ($\Pi= 0$, $\eta\ne0$). The quantities
$I_{0,2}(q,\sigma,z) = 
I_{0,2}(0,\sigma,z) - q^2I_{0,2}^{(2)}(\sigma,z) +{\cal O}(q^4)$ are
defined in the appendix. Let us make a few remarks.
\begin{itemize}
\item In the ``physical'' phase, where parity and flavor symmetries
  are realized in the vacuum, the three pions and the $\eta$ are
  degenerate. The pion degeneracy follows from flavor symmetry, but
  the fact that the $\eta$ is degenerate with the three pions is
  probably an artifact of the large $N$-mean field approximation. The
  three pions and the $\eta$ on the other hand become massless on the
  phase boundary.
\item In the standard Aoki phase $(G_1>G_2)$ the ``charged'' pions are
  massless and the neutral pion and the $\eta$ are massive and
  degenerate, except at the phase boundary where all of them become
  massless.
\item In the non standard Aoki phase $(G_1<G_2)$ the three pions and
  the $\eta$ are massive and degenerate, and all of them become
  massless on the phase boundary.
\item In the special case $G_1 = G_2$, the $\eta$ and the neutral pion
  are massive and degenerate in any of the two vacua, and again become
  massless at the phase boundary.
\end{itemize}


\section{Conclusions}

With the aim of establishing the range of applicability of the
standard $QCD$ chiral effective Lagrangian approach, we have analyzed
in this paper the vacuum structure of a generalized
Nambu--Jona-Lasinio model ($NJL$) with Wilson fermions in the mean
field or leading order $1/N$-expansion. The model has been chosen to
possess the more general $SU(2)_V \times SU(2)_A$ symmetry in the
continuum. The election of the $NJL$ model for
our analysis was motivated by the fact that four-dimensional models
without gauge fields, and with four fermion interactions, are
considered as effective models to describe the low energy physics of
$QCD$.

This generalized $NJL$ model shows a rich phase structure in the 
three-parameter space (two four fermion couplings plus the bare fermion
mass). Indeed, in addition to the standard ``physical'' phases in
which both, parity and flavor symmetries are realized in the vacuum,
new phases where flavor and/or parity are spontaneously broken are
found. These phases are, for $G_1 > G_2$, of the standard Aoki type
with $\la i\bar\psi\gamma_5\tau_3\psi \ra \ne 0$, and $\la
i\bar\psi\gamma_5\psi \ra  = 0$, in agreement also with predictions of
the chiral effective Lagrangian approach. However when $G_1 < G_2$ the
new phase is characterized by $\la i\bar\psi\gamma_5\tau_3\psi \ra =
0$ and $\la i\bar\psi\gamma_5\psi \ra  \ne 0$. In the special case
$G_1 = G_2$, where the continuum $NJL$ action has, in the chiral
limit, a $U(2)_V \times U(2)_A$ symmetry, as $QCD$, two degenerate
vacua, not connected by parity-flavor symmetry transformations, with
broken parity coexist, one with broken and another one with unbroken
flavour symmetry.  

These results show how the phase structure of an
effective model for low energy $QCD$ cannot be entirely understood
from Wilson Chiral Perturbation Theory, based on the standard $QCD$
chiral effective Lagrangian approach. Nonetheless, and as explained in
\cite{monos,POS1,POS2,POS3,monos4}, our conclusions leave no practical
consequences for lattice practitioners: Wilson Chiral Perturbation Theory
is still applicable outside the Aoki phase, and the properties of the
``physical'' phase of lattice QCD with Wilson fermions remain unchanged.

We have also shown in section 4 that the mean field equations of our
generalized $NJL$ model can be obtained
in the leading order of the $1/N$-expansion of a four-fermion model with
non trivial color and flavor interactions, and the same $NJL$-symmetries, 
but where the action is local and free from the sign problem, thus allowing 
to compute corrections to the $1/N$-expansion and to perform numerical
simulations at finite $N$ in order to test the stability of the mean-field
results reported here.

\section*{Acknowledgments}

This work was funded by an INFN-MICINN
collaboration (under grant AIC-D-2011-0663), MICINN (under grants
FPA2009-09638 and FPA2012-35453), DGIID-DGA (grant 2007-E24/2), 
CPAN (Consolider CSD2007-00042), and by
the EU under ITN-STRONGnet (PITN-GA-2009-238353). E.~Follana is
supported on the MICINN Ram\'on y Cajal program, M.~Giordano is
supported by the Hungarian Academy of Sciences under ``Lend\"ulet''
grant No.~LP2011-011, and A.~Vaquero is supported by the Research
Promotion Foundation (RPF) of Cyprus under grant
$\Pi$PO$\Sigma$E$\Lambda$KY$\Sigma$H/NEO$\Sigma$/0609/16.

\appendix

\section{Absence of a sign problem in the infinite-range model}

In this appendix, we prove that the fermion determinant appearing in
Eq.~\eqref{eq:partfunc2} is real and positive. To show this, one  
notices first that $\gamma_4 \Delta_{xy} \gamma_4 = \Delta_{x_P\,y_P}$, 
with $x=(\vec x,x_4)$ and $x_P=(-\vec x,x_4)$. Since the determinants
of $M$ and $M^{(P)}$ are equal, with $M^{(P)}_{xy}= M_{x_P y_P}$, one finds that
\begin{equation}
  \label{eq:sign1}
  \begin{aligned}
\Det M = \Det M^{(P)} &= \Det[\gamma_4\Delta\gamma_4 +
  \sigma  + i\gamma_5\vec\tau \cdot\vec\pi  +  i\gamma_5 \eta +
  \vec\tau\cdot\vec\rho] \\ &= \Det[\Delta +
  (\sigma  - i\gamma_5\vec\tau \cdot\vec\pi  -  i\gamma_5 \eta +
  \vec\tau\cdot\vec\rho\,)]\,.
  \end{aligned}
\end{equation}
Next, one exploits the unitarity of the matrix $\tilde C=\gamma^1\gamma^3$ 
and the fact that $\tilde C \Delta \tilde C^\dag = \Delta^*$ 
to write
\begin{equation}
  \label{eq:sign2}
  \Det M = \Det[\Delta^* +
  (\sigma  - i\gamma_5\vec\tau \cdot\vec\pi  -  i\gamma_5 \eta +
  \vec\tau\cdot\vec\rho)]\,.
\end{equation}
Finally, one performs a rotation $R$ in flavor space in such a way
that the vectors $\vec\pi_R = R\vec\pi$ and $\vec\rho_R = R\vec\rho$
lie in the (1,3)-plane, i.e., so that $\vec\rho_R\wedge\vec\pi_R$ is
along direction 2 in flavor space. Since $R$ is implemented by a
unitary operator $U_R$, and $\tau_1$ and $\tau_3$ are real, one has that
\begin{equation}
  \label{eq:sign3}
  \begin{aligned}
  \Det M &= \Det[\Delta^* +
  (\sigma  - i\gamma_5\vec\tau \cdot\vec\pi_R  -  i\gamma_5 \eta +
  \vec\tau\cdot\vec\rho_R)] \\ &=
\Det[\Delta +
  (\sigma +  i\gamma_5\vec\tau \cdot\vec\pi_R  +  i\gamma_5 \eta +
  \vec\tau\cdot\vec\rho_R)]^*
\\ &= \Det[\Delta +
  (\sigma +  i\gamma_5\vec\tau \cdot\vec\pi  +  i\gamma_5 \eta +
  \vec\tau\cdot\vec\rho)]^* = \Det M^*
\,.
  \end{aligned}
\end{equation}
Finally, since $M$ is diagonal and trivial in color, $M=\bar M {\bf
  1}_C$, one has that for an even number of colors $\det M = 
(\det \bar M)^N = (\det \bar M \det \bar M^\dag)^{\f{N}{2}} \ge 0$.

\section{Quadratic part of the action in the local model}

In this appendix we report a few results related to the local model
discussed in Section \ref{sec:4}. We use the following notation:
\begin{equation}
  \label{eq:loc_14}
  \begin{aligned}
    I_0(q,\sigma,z) &=  \int_B\f{d^4p}{(2\pi)^4}
  \f{1}{g(p+\f{q}{2},z)g(p-\f{q}{2},z)}\bigg\{
    \sum_{\mu=1}^4\sin(p_\mu+\textstyle\f{q_\mu}{2})
    \sin(p_\mu-\textstyle\f{q_\mu}{2}) + \\ &\phantom{bababababababa}
    [w_r(p+\textstyle\f{q}{2}) + m_0+\sigma]
    [w_r(p-\textstyle\f{q}{2}) + m_0+\sigma] + z^2\bigg\}\,,\\
    I_1(q,\sigma,z) &= \int_B\f{d^4p}{(2\pi)^4}
  \f{[w_r(p+\textstyle\f{q}{2}) + m_0+\sigma]
    [w_r(p-\textstyle\f{q}{2}) + m_0+\sigma]}{g(p+\f{q}{2},z)g(p-\f{q}{2},z)}
     \,,\\
     I_2(q,\sigma,z) &= \int_B\f{d^4p}{(2\pi)^4}
     \f{1}{g(p+\f{q}{2},z)g(p-\f{q}{2},z)}
     \,,\\
     I_3(q,\sigma,z) &= \int_B\f{d^4p}{(2\pi)^4}
  \f{w_r(p+\textstyle\f{q}{2}) + w_r(p-\textstyle\f{q}{2})+ 2(m_0+\sigma)
    }{g(p+\f{q}{2},z)g(p-\f{q}{2},z)}
     \,.
  \end{aligned}
\end{equation}
The non-vanishing entries of the inverse propagator $G^{-1}_{ij}(q)$,
Eq.~\eqref{eq:loc_11}, are listed below for the standard-Aoki and non
standard-Aoki cases. In the unbroken phase the inverse propagator is
diagonal, and can be recovered by simply setting $\Pi=0$ or $\eta=0$
in the equations below. Here $\Pi=|\vec\pi|$. 
\begin{equation}
  \label{eq:local_inv_prop}
  \begin{aligned} 
&    \text{Standard Aoki phase:} \\
&G^{-1}_{\sigma\sigma}(q) = 8\left\{\f{\beta_1}{4} - I_0(q,\sigma,\Pi) + 2
      I_1(q,\sigma,\Pi)\right\}\,,\\
&    G^{-1}_{\pi_a\pi_b}(q) =8\left\{\left[\f{\beta_1}{4} -
        I_0(q,\sigma,\Pi)\right]\delta_{ab} + 2\pi_a\pi_b I_2(q,\sigma,\Pi)\right\}
    \,,\\
&    G^{-1}_{\eta\eta}(q)= 8\left\{\f{\beta_2}{4} -
        I_0(q,\sigma,\Pi) + 2\Pi^2 I_2(q,\sigma,\Pi)\right\}\,,\\
&    G^{-1}_{\rho_a\rho_b}(q)= 8\left\{\left[\f{\beta_2}{4} -
        I_0(q,\sigma,\Pi) + 2 I_1(q,\sigma,\Pi)\right]\delta_{ab}+ 2
        I_2(q,\sigma,\Pi)(\Pi^2\delta_{ab}-\pi_a\pi_b)
\right\}
\,,\\  
&      G^{-1}_{\sigma\pi_a}(q) = -8I_3(q,\sigma,\Pi)\pi_a\,,\\
&      G^{-1}_{\eta\rho_a}(q) = -8I_3(q,\sigma,\Pi)\pi_a\,.\\ 
&\text{Non Standard Aoki phase:} \\
& G^{-1}_{\sigma\sigma}(q) = 8\left\{\f{\beta_1}{4} - I_0(q,\sigma,\eta) + 2
      I_1(q,\sigma,\eta)\right\}\,,\\
&    G^{-1}_{\pi_a\pi_b}(q) =8\left\{\f{\beta_1}{4} -
        I_0(q,\sigma,\eta)+ 2\eta^2 I_2(q,\sigma,\eta)\right\}\delta_{ab}
    \,,\\
&    G^{-1}_{\eta\eta}(q)= 8\left\{\f{\beta_2}{4} -
        I_0(q,\sigma,\eta) + 2\eta^2 I_2(q,\sigma,\eta)\right\}\,,\\
&    G^{-1}_{\rho_a\rho_b}(q)= 8\left\{\f{\beta_2}{4} -
        I_0(q,\sigma,\eta) + 2 I_1(q,\sigma,\eta)\right\}\delta_{ab}\,,\\  
&      G^{-1}_{\sigma\eta}(q) = -8I_3(q,\sigma,\eta)\eta\,,\\
&      G^{-1}_{\pi_a\rho_b}(q) = -8I_3(q,\sigma,\eta)\eta\delta_{ab}\,.
  \end{aligned}
\end{equation}
The first two terms in the small-$q$ expansion of $I_0$ and $I_2$ is
given below. 
\begin{equation}
  \label{eq:small_q}
  \begin{aligned}
    I_0(q,\sigma,z) &=   \int_B\f{d^4p}{(2\pi)^4}
    \f{1}{g(p,z)} \\ &\phantom{diom} -\f{q^2}{8}\int_B\f{d^4p}{(2\pi)^4}
    \f{\sum_{\mu=1}^4 (\cos p_\mu)^2 + r^2(\sin p_\mu)^2}{[g(p,z)]^2} +
    {\cal O}(q^4) \\
    & = I_0(0,\sigma,z) - q^2I_0^{(2)}(\sigma,z) +{\cal O}(q^4)\,, \\
    I_2(q,\sigma,z) &= \int_B\f{d^4p}{(2\pi)^4}
    \f{1}{[g(p,z)]^2}  \\ &\phantom{diom}
    -\f{q^2}{2}\int_B\f{d^4p}{(2\pi)^4}\f{\sum_{\mu=1}^4 (\sin
      p_\mu)^2[\cos p_\mu + r(w_r(p)+m_0+\sigma)]^2}{[g(p,z)]^4} + 
    {\cal O}(q^4)\\
    & = I_2(0,\sigma,z) - q^2I_2^{(2)}(\sigma,z) +{\cal O}(q^4)\,.
  \end{aligned}
\end{equation}
Notice that
\begin{equation}
  \label{eq:small_q2}
I_0(0,\sigma,z) =    \int_B\f{d^4p}{(2\pi)^4}
    \f{1}{g(p,z)} = \left\{
  \begin{aligned}
    &\f{\beta_1}{4}\,, &&& &\text{in the standard Aoki phase,}\\
    &\f{\beta_2}{4}\,, &&& &\text{in the non standard Aoki phase.}
  \end{aligned}\right.
\end{equation}


\begin{thebibliography}{99}

\bibitem{an}
V. Azcoiti, A. Nakamura,
Phys. Rev. {\bf D27}, 2559 (1983).

\bibitem{hamber}
H.W. Hamber,
Nucl. Phys. {\bf B251}, 182 (1985).

\bibitem{a1}
S. Aoki,
Phys. Rev. {\bf D30}, 2653 (1984).

\bibitem{a2}
S. Aoki,
Phys. Rev. Lett. {\bf 57}, 3136 (1986).

\bibitem{aokinjl}
S. Aoki, S. Boettcher, A. Gocksch, Phys. Lett. {\bf B331}, 157 (1994).


\bibitem{creutz}
M. Creutz, arXiv:hep-lat/9608024.

\bibitem{bit}
K.M. Bitar,
Phys. Rev. {\bf D56}, 2736 (1997).

\bibitem{aku}
S. Aoki, T. Kaneda, and A. Ukawa,
Phys. Rev. {\bf D56}, 1808 (1997).

\bibitem{heller}
K.M. Bitar, U.M. Heller, and R. Narayanan,
Phys. Lett. {\bf B418}, 167 (1998).

\bibitem{heller2}
R.G. Edwards, U.M. Heller, R. Narayanan, and R.L. Singleton, Jr.,
Nucl. Phys. {\bf B518}, 319 (1998).

\bibitem{sharpe}
S.R. Sharpe, R.L. Singleton, Jr,
Phys. Rev. {\bf D58}, 074501 (1998).

\bibitem{a3}
S. Aoki,
Nucl. Phys. {\bf B} (Proc. Suppl.) {\bf 60A}, 206 (1998).

\bibitem{bit2}
K. Bitar,
Nucl. Phys. {\bf B} (Proc. Suppl.) {\bf 63A-C}, 829 (1998).

\bibitem{shasi}
S. Sharpe, R.L. Singleton Jr,
Nucl. Phys. {\bf B} (Proc. Suppl) {\bf 73}, 234 (1999).

\bibitem{kps}
R. Kenna, C. Pinto, and J.C. Sexton,
Phys. Lett. {\bf B505}, 125 (2001).

\bibitem{gosha}
M. Golterman, Y. Shamir,
Phys. Rev. {\bf D68}, 074501 (2003).

\bibitem{ilg}
E.M. Ilgenfritz, W. Kerler, M. M\"uller-Preussker, A. Sternbeck, and
H. Stuben,
Phys. Rev. {\bf D69}, 074511 (2004).

\bibitem{ster}
A. Sternbeck, E.M. Ilgenfritz, W. Kerler, M. M\"uller-Preussker, and
H. Stuben,
Nucl. Phys. {\bf B} (Proc. Suppl.) {\bf 129\&130}, 898 (2004).

\bibitem{golt}
M. Golterman, S.R. Sharpe, and R.L. Singleton, Jr,
Nucl. Phys. {\bf B} (Proc. Suppl.) {\bf 140}, 335 (2005).


\bibitem{ilg3}
E.M. Ilgenfritz, M. M\"uller-Preussker, M. Petschlies, K. Jansen,
M.P. Lombardo, O. Philipsen, L.Zeidlewicz, and A. Sternbeck,
{\sl PoS} {\bf Lattice2007}, 238 (2008).


\bibitem{monos}
V. Azcoiti, G. Di Carlo, A. Vaquero,
Phys. Rev. {\bf D79}, 014509 (2009).

\bibitem{sharpe2}
S. Sharpe, Phys. Rev. {\bf D79},  054503 (2009).

\bibitem{POS1}
A. Vaquero, V. Azcoiti, G. Di Carlo, E. Follana,
{\sl PoS} {\bf LATTICE2009} 068 (2009).

\bibitem{POS2}
V. Azcoiti, G. Di Carlo, E. Follana, A. Vaquero,
{\sl PoS} {\bf LATTICE2010} 091 (2010).

\bibitem{POS3}
V. Azcoiti, G. Di Carlo, E. Follana, A. Vaquero,
{\sl PoS} {\bf LATTICE2011} 112 (2011).

\bibitem{verba}
G. Akemann, P.H. Damgaard, K. Splittorff, J.J.M. Verbaarschot, 
Phys. Rev. {\bf D83}, 085014 (2011).

\bibitem{Splittorff}
M. Kieburg, K. Splittorff, J.J.M.  Verbaarschot,
Phys. Rev. {\bf D85} 094011 (2012).

\bibitem{witten}
C. Vafa, E. Witten,
Phys. Rev. Lett. {\bf 53}, 535 (1984).

\bibitem{monos2}
V. Azcoiti, A. Galante,
Phys. Rev. Lett. {\bf 83}, 1518 (1999).

\bibitem{ji}
X. Ji,
Phys. Lett. {\bf B554}, 33 (2003).

\bibitem{kr}
P.R. Crompton,
Phys. Rev. {\bf D72}, 076003 (2005).

\bibitem{monos3}
V. Azcoiti, G. Di Carlo, E. Follana, A. Vaquero,
JHEP{\bf 1007:}047, (2010).

\bibitem{bitar} K. M. Bitar, P. M. Vranas, Phys.Rev. {\bf D50}, 3406 (1994).


\bibitem{monos4}
V. Azcoiti, G. Di Carlo, E. Follana, A. Vaquero, Nucl. Phys. {\bf
  B870}, 138 (2013). 

\bibitem{shankar}
A. Dhar, R. Shankar, and S. R. Wadia,
Phys. Rev. {\bf D31}, 3256 (1985).

\bibitem{ref1}
D. Ebert, M. Nagy, M.K. Volkov,
Phys. Atom. Nucl. {\bf 59}, 140 (1996);
Yad. Fiz. {\bf 59}, 149 (1996).

\bibitem{ref2}
D. Ebert, H. Reinhardt, M.K. Volkov,
Prog. Part. Nucl. Phys. {\bf 33}, 1 (1994).












\end{thebibliography}
\end{document}